\documentclass[12pt]{article}
\usepackage{amsmath}
\usepackage{amssymb} 
\usepackage{bbm} 
\usepackage[small]{caption2} 
\usepackage{fleqn} 
\usepackage{graphicx} 
\usepackage{mathrsfs}	
\usepackage[small,loose]{subfigure}  
\usepackage{nomencl} 
\usepackage[hypertex]{hyperref}
\setcaptionwidth{.85\textwidth}
\advance \headheight by 3.0truept       

\DeclareMathOperator{\im}{Im}

\newcommand{\CenterObject}[1]{\ensuremath{\vcenter{\hbox{#1}}}}
\newcommand{\D}{\mathrm{d}}
\newcommand{\I}{\mathrm{i}}

\begin{document}
\thispagestyle{empty}
\flushright
IPMU 08-0041\\
UCB-PTH-08/62\\
LA-UR-08-04779\\
TUM-HEP 08/691\\
\vfil
\begin{center}
{\Large\bf Unified origin of baryons and dark matter}\\[2cm]

{\large Ryuichiro Kitano$^{1}$, Hitoshi Murayama$^{2,3,4}$, Michael
Ratz$^{5}$ } \\[1cm] {\it
${}^1$ Theoretical Division T-8, Los Alamos National Laboratory,
Los Alamos, NM 87545\\[0.2cm]
${}^2$ Institute for the Physics and Mathematics of the Universe,\\
University of Tokyo, Kashiwa, Chiba 277-8568, Japan\\[0.2cm]
${}^3$ Department of Physics,
University of California, Berkeley, CA 94720\\[0.2cm]
${}^4$ Theoretical Physics Group, Lawrence Berkeley National
Laboratory,
Berkeley, CA 94720\\[0.2cm]
${}^5$ Physik Department T30, Technische Universit\"at M\"unchen,\\
James-Franck-Stra\ss{}e, 85748 Garching, Germany\\[0.2cm]
}
\end{center}

\abstract{We investigate the possibility that both the baryon asymmetry
of the universe and the observed cold dark matter density are generated
by decays of a heavy scalar field which dominates the universe before
nucleosynthesis. Since baryons and cold dark matter have common origin,
this mechanism yields a natural explanation of the similarity of the
corresponding energy densities. The cosmological moduli and gravitino
problems are avoided.}  \vfil \clearpage

\section{Introduction}

Supersymmetry is a well-motivated framework for TeV scale physics and it has
several conceptually nice features. Apart from being the unique extension of the
space-time symmetry it provides us with the most compelling scenarios of gauge
coupling unification.
Not only for short-distance physics, supersymmetry provides interesting
possibilities for the explanation of the structure of the Universe. The
supersymmetric Standard Model contains a natural candidate for cold dark
matter of the Universe, and also the fact that quadratic divergences are
absent allows us to naturally postpone the cut-off of the theory as high
as the Planck scale. It makes us possible to calculate high-scale or
high-temperature phenomena such as inflation and reheating in a reliable
framework.

However, a closer inspection of cosmological scenarios for dark matter
production and baryogenesis in supersymmetric models reveals that there is often
an inconsistency in the underlying assumptions.
For example, the most popular scenarios for dark matter and for
baryogenesis are known to be incompatible. It is widely accepted that
the relic density of the neutralino from thermal decoupling naturally
explains the amount of cold dark matter, and there is a good explanation
of the baryon component of the Universe by the thermal leptogenesis
scenario~\cite{Fukugita:1986hr}. The thermal neutralino dark matter
realizes in the gravity mediation scenario, i.e., the gravitino mass
$m_{3/2} \sim 100$~GeV, whereas the thermal leptogenesis needs a high
reheating temperature after inflation, $T_R \gtrsim
10^{9}$~GeV~\cite{Davidson:2002qv,Giudice:2003jh,Buchmuller:2004nz}.
Considering the thermal production of the gravitinos and their decays,
the bound from the big bang nucleosynthesis (BBN) excludes such a
possibility~(see e.g.\
\cite{Pagels:1981ke,Weinberg:1982zq,Khlopov:1984pf,Bolz:2000fu,Kawasaki:2008qe}).
It is possible to avoid the constraints from the BBN by assuming a very
heavy ($m_{3/2} \gtrsim 100$~TeV) or a stable gravitino, but the
most popular explanation of dark matter via the
thermal relic neutralinos will be lost in such cases since the lightest
supersymmetric particle will be the $W$ino~\cite{Randall:1998uk,Giudice:1998xp}
(or Higgsino~\cite{Choi:2005ge}),
which typically leads to a too small abundance, or the gravitino.

Not only that, for $m_{3/2} \sim 100$~GeV, there is a serious cosmological
moduli (or Polonyi) problem~\cite{Coughlan:1983ci}. It is possible to give a
large mass to the Polonyi field by assuming a non-trivial interactions in order
to avoid the late time decay~\cite{Dine:1983ys}, but in such a case the decay
into gravitinos causes a disaster~\cite{Joichi:1994ce,Ibe:2006am}.
There is also a rather disturbing bound on the reheating temperature
coming from modular cosmology~\cite{Buchmuller:2004xr}.

It is clear from the above discussion that we need to abandon some of
the ``standard'' assumptions. We propose in this paper an extreme but
cosmologically safe scenario in which radiation, dark matter and baryon
asymmetry are all generated right before the BBN era so that there is no
constraint from the high-temperature gravitino production, moduli
dynamics etc.\
(See~\cite{Dimopoulos:1987rk,Cline:1990bw,Thomas:1995ze,Fujii:2002aj,Barenboim:2008zk}
for earlier works). We find that there is a consistent framework to
realize this scenario by a late-time decay of a scalar condensate $\phi$
in supersymmetric models.
We assume that coherent oscillations of a scalar field $\phi$ first dominate the
energy density of the universe. The $\phi$ field then decays later producing a
large amount of radiation and all the dangerous relics (e.g.\ gravitinos) get
diluted to a negligible density. The standard BBN takes place after $\phi$ has
decayed at sufficiently high temperature. The dark matter is produced by the
$\phi$ decay and the correct abundance is obtained in the $W$ino/Higgsino LSP
scenario due to pair annihilation~\cite{Moroi:1999zb} which is less effective
than in the so-called thermal scenarios. The baryon asymmetry is also generated
in $\phi$ decays.
The decay channel of a $\phi$ field into two gravitinos is forbidden by 
$R$-parity, and we assume that the decay into a gravitino and the fermionic
superpartner of $\phi$ $(\widetilde{\phi})$ is kinematically forbidden.
Interestingly, by unifying the origin of dark matter and baryon
asymmetry, this scenario explains the one of the puzzling issues of our
Universe that the energy densities of dark matter and baryon asymmetry
are close to each other, $\Omega_b \sim \Omega_\mathrm{CDM}$.\footnote{
See, for example,
\cite{Dodelson:1991iv,Kaplan:1991ah,Kuzmin:1996he,Laine:1998rg,Fujii:2002aj,Kitano:2004sv}
for earlier attempts to explain the similarity: $\Omega_b \sim
\Omega_\mathrm{CDM}$.  }

The paper is organized as follows:
in Section~\ref{sec:Baryogenesis} we present the baryogenesis mechanism, and
show that it is indeed possible to generate the observed amount of baryon
asymmetry while satisfying the BBN constraints on the reheating temperature
after the $\phi$ decay.
In Section~\ref{sec:DarkMatter}, we discuss the abundance of the dark matter
from the $\phi$ decay.
In particular, in our preferred scenario the ratio $\Omega_\mathrm{CDM}/\Omega_b
\sim 5$ implies a large gravitino mass, $m_{3/2} \sim
100\,\mathrm{TeV}$ which fits nicely to the $W$ino/Higgsino LSP scenario.

\section{Baryogenesis}
\label{sec:Baryogenesis}

\subsection{Basic idea}
We consider a chiral superfield $\Phi=(\phi,\widetilde{\phi},F_\phi)$ which
couples to the matter fields via a higher-dimensional term in the
superpotential \cite{Thomas:1995ze},
\begin{equation}\label{eq:SuperpotentialPhiUDD}
 \mathscr{W}\,\supset\, 
 \frac{1}{M}\Phi\,U\,D\,D\;,
\end{equation}
with $U=(\widetilde{u}^c,u^c,F_u)$ and $D=(\widetilde{d}^c,d^c,F_d)$ denoting
the up- and down-type quark superfields. Here, we suppressed color and
generation indices, and absorbed dimensionless couplings into $M$ which will be
taken to be of the of order of the Planck scale,  $M\sim
M_\mathrm{P}=2.44\times10^{18}\,\mathrm{GeV}$, unless stated otherwise. 
Due to this operator $\phi$ effectively carries baryon number (+1).

Let us now define the $\phi$ number asymmetry 
\begin{equation}\label{eq:qphi}
 q_\phi\,:=\,\I\left(\Dot{\phi}^*\phi-\phi^*\Dot{\phi}\right)\;.
\end{equation}
$q_\phi$ is given by the difference between the number densities $n_\phi$ and
$n_{\phi^*}$ of particles $\phi$ and antiparticles $\phi^*$. $q_\phi$ can be
interpreted as  angular momentum of the $\phi$ field rotating in the complex
plane \cite{Affleck:1985fy}.

The scenario we shall describe in the following consist of the following
sequence of steps: first, a positive $q_\phi$ is generated. Then there is an
era of coherent $\phi$ oscillations where a significant fraction of the energy
density of the universe is carried by these oscillations, and $q_\phi$ is
conserved. 
Finally, $\phi$ number is converted into the baryon number by its decay,
and it reheats the universe up to $O(100\,\mathrm{MeV})$ consistently with
nucleosynthesis.

Before we describe the mechanism in detail, let us briefly explain the
main differences to Ref.~\cite{Thomas:1995ze}. For the mechanism to
work, one has make sure that dangerous $\phi$-number violating
interaction terms are absent or sufficiently suppressed.  One can forbid
these terms by imposing a symmetry. In Ref.~\cite{Thomas:1995ze}, a
model with a $\mathbbm{Z}_{4\,\mathrm{R}}$ symmetry is presented, which
ensures the $\phi$-number conservation at a sufficient level and also
prevents $\phi$ from dominating the universe. Below, we will consider a
different model with an anomaly-free $\mathbbm{Z}_{9}$ discrete baryon
symmetry in addition to the usual $R$-parity (see
Table~\ref{tab:Z6RChargeAssignments}), and will, as already stated,
assume that that $\phi$ dominates the universe at an early epoch.

In order to obtain the baryon asymmetry before BBN, in Ref.~\cite{Thomas:1995ze}
enhanced couplings of $\phi$ to the baryons are assumed such that the $\phi$
lifetime is short enough. Moreover, in the case of the $\phi$ mass of the order
$100\,\mathrm{GeV}$, the universe would be always matter ($\phi$ or dark matter)
dominated once $\phi$ dominates the universe at an early time.
This is not compatible with the requirement of successful BBN and, therefore, a
model without (early) $\phi$ domination is constructed there.

%
The situation is, however, different if $\phi$ is heavy. As the temperature
after the $\phi$ decay is higher than the BBN temperature, $\phi$ is allowed to
dominate the energy density of the universe.  This high temperature after $\phi$
decay also plays a crucial role in the generation of cold dark matter (cf.\
Section~\ref{sec:DarkMatter}).

\subsection{$\boldsymbol{\phi}$ evolution}

Let us start by considering the dynamics of the $\phi$ field.
The evolution of $\phi$ is described by its equation of motion,
\begin{equation}\label{eq:EOMphi}
 \Ddot{\phi}+(3H+\Gamma_\phi)\,\Dot{\phi}+\frac{\partial V}{\partial \phi^*}
 ~=~0\;,
\end{equation}
where $V=V(\phi,\phi^*,\dots)$ denotes the scalar potential, $H$ the Hubble
rate and $\Gamma_\phi$ the $\phi$ decay rate.
Eq.~\eqref{eq:EOMphi} translates into an equation of motion for $q_\phi$,
\begin{equation}\label{eq:EOMqphi}
 \Dot{q}_\phi+3H\,q_\phi
 ~=~
 -\I\left(\phi\frac{\partial V}{\partial \phi}-
 \phi^*\frac{\partial V}{\partial \phi^*}\right)
 \;.
\end{equation}
Hence, a non-vanishing right-hand side of \eqref{eq:EOMqphi} can be used for the
first step, i.e.\ to create non-zero $q_\phi$ dynamically.  Before explaining
this in detail, recall that we need also to satisfy the condition of $\phi$ 
number conservation in the stage of $\phi$ oscillation. This means that in the
$\phi$ oscillation era the $\phi$ number violating terms have to be absent (or
sufficiently suppressed). The most dangerous term of this type is
$\mu^2\phi^2+\text{h.c.}$.

In order to enforce the absence of those dangerous terms, we impose a discrete
$\mathbbm{Z}_{9}$ symmetry which is an anomaly free subgroup of baryon number
symmetry~\cite{Ibanez:1991pr,Ibanez:1991zv,Kurosawa:2001iq,Babu:2003qh,Davoudiasl:2005ks}.
The charge assignment is listed in Tab.~\ref{tab:Z6RChargeAssignments}, where
$\bar{\Phi}$ is introduced in order to give a mass term for the fermionic
superpartner of $\phi$ without introducing $\mu^2 \phi^2$ term in the
Lagrangian.\footnote{The analysis will remain unchanged when we include the
dynamics of the $\bar \phi$ field. Although a possible mass mixing term, $m^2
\phi \bar \phi + \text{h.c.}$, will distribute the baryon number to the $\bar
\phi$ field, Eq.~\eqref{eq:EOMqphi} will be the same once we include the $\bar
\phi$ field in $q_\phi$ in Eq.~\eqref{eq:qphi}. The decay of $\bar \phi$ (or
more precisely the other mass eigenstate) happens about the same time as the
$\phi$ decay provided the mixing is order one.} With this choice $\phi=0$ can
always be a minimum of the potential which is necessary to preserve the
$R$-parity.
\begin{table}[t]
\begin{center}
$
\begin{array}{c|cccccccccc}
 \text{field}  & Q & U & D & L & E & N & H  & \bar H & \Phi & \bar{\Phi} \\
 \hline
 \hline
 \mathbbm{Z}_{9} & +1 & -1 & -1  &  0 &  0 & 0  & 0 & 0 & +3 & -3 \\
 R\text{-parity} & - & - & -  &  - &  - & -  & + & + & - & - \\
\end{array}
$
\end{center}
 \caption{Charge assignments under $\mathbbm{Z}_{9}$ and $R$-parity.
 We denote the MSSM superfields according to $Q=(q,\widetilde{q},F_q)$ 
with $q$ representing the quark doublets
 etc.}
 \label{tab:Z6RChargeAssignments}
\end{table}

The symmetry does, however, allow for a $\Phi^6$ term in the K{\" a}hler
potential and in the superpotential.  In the following we discuss the
two cases in which the $\Phi^6$ term in superpotential is absent and
present.

\subsubsection*{Case A: no $\boldsymbol{\Phi^6}$ term in the superpotential}
In the case where there is a $\Phi^6$ term only in the K{\" a}hler potential,
the potential for the $\phi$ field is given by
\begin{eqnarray}\label{eq:Vphi6}
 V
 &=&
 m_\phi^2 | \phi |^2 
+  m_{3/2}^2 M^2 F(| \phi |^2 / M^2 )
\nonumber \\
&&
+ \left[
\kappa\,\frac{m_{3/2}^2}{M^4}\,\phi^6+\text{h.c.}
\right]
 +\text{higher-order terms}\;,
\end{eqnarray}
where $\kappa$ is expected to be order one. The gravitino mass parameter
$m_{3/2}$ represents the supersymmetry breaking scale. The function
$F(x)$ is a general (polynomial) function.

The presence of the $\phi^6$ term  in \eqref{eq:Vphi6} can lead to a dynamical
generation of $q_\phi\ne0$ as follows:  for $H \gg m_\phi$, $\phi$ oscillation
is negligible and  we can treat $\phi$ to be constant. We can integrate the
equation
\begin{equation}
 \Dot{q}_\phi + 3 H\, q_\phi 
 ~=~ 
 \frac{m_{3/2}^2}{M^4}\,\im\left[ \kappa\,\phi^6\right]\;,
 \label{eq:diffeq}
\end{equation}
so that
\begin{equation}\label{eq:qphi2}
 q_\phi( t = m_\phi^{-1} )
 ~\sim~
 |\kappa|\,
 \frac{m_{3/2}^2}{2m_\phi\,M^4}\,\phi_\mathrm{ini}^6
 \;,
\end{equation}
where $\phi_\mathrm{ini}$ is the initial amplitude of $\phi$ after inflation
which is generically $O(M)$ with the potential in Eq.~\eqref{eq:Vphi6}
(with $m_{3/2}^2$ replaced by $O(H^2)$). The
number density of $\phi$ and $\phi^*$ particles is given by $\rho_\phi/m_\phi$
where $\rho_\phi\simeq m_\phi^2\,|\phi|^2+|\dot\phi|^2$. Since the oscillation
of $\phi$ starts with amplitude $\phi_\mathrm{ini}$, we obtain for the
dimensionless $\phi$ asymmetry
\begin{equation}\label{eq:epsilon}
 \varepsilon
 ~:=~
 \frac{q_\phi}{n_\phi+n_{\phi^*}}
 ~\sim~
 |\kappa|\,\left(\frac{m_{3/2}}{m_\phi}\right)^2 \;.
\end{equation}
Here we have taken $\phi_\mathrm{ini} \sim M$. We have checked that this
expression yields roughly the correct order of magnitude for $\varepsilon$ (as
long as $\phi_\mathrm{ini}$ is comparable to $M$) by solving the equation of
motion numerically.

The resulting $\phi$ asymmetry stays constant after $\phi$ starts to oscillate
because of the r.h.s.\ of Eq.~\eqref{eq:diffeq} becomes numerically irrelevant
when the amplitude drops (far) below $M$.
As a consequence, $R^3q_\phi$ (with $R$ being the scale factor) is approximately
conserved during the $\phi$ oscillation era, until $\phi$ decays.

\subsubsection*{Case B: $\boldsymbol{\Phi^6}$ term in the superpotential}

The case with $\Phi^6$ term in the superpotential is qualitatively the same as
case A.
The potential of the $\phi$ field in this case is given by
\begin{eqnarray}
\label{eq:Vphi6-2}
 V
 &=&
 m_\phi^2 | \phi |^2 
+  m_{3/2}^2 M^2 F(| \phi |^2 / M^2 )
\nonumber \\
&&
+ \left[
\kappa^\prime\,\frac{m_{3/2}}{M^3}\,\phi^6+\text{h.c.}
\right]
+ \kappa^{\prime \prime}\,\frac{|\phi|^{10}}{M^6}
 +\text{higher-order terms}\;,
\end{eqnarray}
where $\kappa^\prime$ and $\kappa^{\prime \prime}$ are $O(1)$ coefficients. In
the early universe, $m_{3/2}$ which represents the SUSY breaking effect is
replaced by the Hubble rate $H$. Now the minimum of the potential is generically
$\phi \sim (HM^3)^{1/4}$ which sets the initial amplitude of the $\phi$
oscillation to be $\phi_\mathrm{ini} \sim (m_\phi M^3)^{1/4}$ since the $\phi$
oscillations start when $H\sim m_\phi$.

The equation for the evolution of the $\phi$ number asymmetry $q_\phi$
in Eq.~(\ref{eq:EOMqphi}) is
\begin{equation}
\Dot{q}_\phi + 3 H\, q_\phi 
~=~ 
\frac{m_{3/2}}{M^3}\,\im\left[ \kappa^\prime\,\phi^6\right]\;.
\label{eq:diffeq-2}
\end{equation}
With the value of $\phi$ at the minimum of the potential, $\phi \sim
(HM^3)^{1/4}$, before $\phi$ oscillation, we find
\begin{eqnarray}
 q_\phi \sim |\kappa^{\prime}| \frac{m_{3/2}}{m_\phi} \frac{\phi_{\rm ini}^6}{M^3}\ .
\end{eqnarray}
Therefore, the asymmetry factor $\varepsilon$ is estimated to be
\begin{eqnarray}
 \varepsilon \sim |\kappa^\prime| \frac{m_{3/2}}{m_\phi}\ ,
\end{eqnarray}
which is larger than the case without the $\Phi^6$ term in the
superpotential if $m_{3/2} \ll m_{\phi}$ and $\kappa^\prime \sim 1$. In
conclusion, the presence of the $\Phi^6$ term leads only to a quantitatively
different result.

\subsection{Baryogenesis via $\boldsymbol{\phi}$ decay}
\label{sec:PhiDecay}

So far we have seen that, due to the presence of higher-order terms, a $\phi$
number asymmetry is induced which is conserved in the regime of $\phi$
oscillations until $\phi$ decays.
Let us now consider the conversion of $\phi$ number to baryon number of the
universe through the decay arising from the coupling
\eqref{eq:SuperpotentialPhiUDD}, $\phi \to q q \widetilde{q}$.
The corresponding decay rate is given by
\begin{equation}
 \Gamma_\phi~=~\xi\,\frac{m_\phi^3}{M^2}\;,
\end{equation}
where $\xi$ is obtained by a standard calculation. In the simplest case where
all couplings of $\phi$ to the quark superfields equal one,
we obtain $\xi=27/(256\pi^3)\simeq 3\times 10^{-3}$.

The temperature $T_d$ of the thermal bath after $\phi$ decay is calculated by 
equating Hubble rate $H$ and $\Gamma_\phi$,
\begin{equation}
 T_d~\simeq~120\,\mathrm{MeV}
\left(\frac{\xi}{10^{-2}}\right)^{1/2}
\left(\frac{m_\phi}{1500\,\mathrm{TeV}}\right)^{3/2}
\left(\frac{M_\mathrm{P}}{M}\right)\;.
\end{equation}
Since $\phi$ has a large hadronic branching fraction, $T_d$ has to fulfill
\cite{Kawasaki:2000en} $T_d\gtrsim4\,\mathrm{MeV}$. The corresponding lower
bound on the $\phi$ mass is 
\begin{equation}
 m_\phi\,\gtrsim\,150\,\mathrm{TeV}\,
\left(\frac{10^{-2}}{\xi}\right)^{1/3}
 \left(\frac{M}{M_\mathrm{P}}\right)^{2/3}\;,
\label{eq:bound-mphi}
\end{equation} 
where we have exploited that $\sqrt{\frac{\pi^2g_*}{90}}\simeq1$ for
temperatures of the order MeV.

Assuming that $\phi$ dominates the energy density of the universe before its
decay, the number density of $\phi$ just before decay is obtained by
\begin{equation}
 m_\phi\,(n_\phi+n_{\phi^*})\,\simeq\,\frac{\pi^2}{30}\,g_*\,T_d^4\;.
\end{equation}
Using the relation between baryon asymmetry and $\phi$ number, 
$n_b=q_{\phi}=\varepsilon (n_\phi+n_{\phi^*})$, we can estimate the baryon
asymmetry as 
\begin{eqnarray}
 \frac{n_b}{s}
 & \simeq & \frac{3}{4}\,\varepsilon\,\frac{T_d}{m_\phi}
 \label{eq:baryon}
 \\
 & \sim &
\left \{
\begin{array}{lr}
{\displaystyle
 10^{-10}
\cdot |\kappa|\,
\left(\frac{\xi}{10^{-2}}\right)^{1/2}\,
\left(\frac{m_{3/2}}{50~\mathrm{TeV}}\right)^2\,
\left(\frac{m_\phi}{1500\,\mathrm{TeV}}\right)^{-3/2}\,
\left(\frac{M_\mathrm{P}}{M}\right)\;,
} & \:{\mbox{(A)}}
 \\[0.2cm]
{\displaystyle 10^{-10}
\cdot |\kappa^\prime|\,
\left(\frac{\xi}{10^{-2}}\right)^{1/2}\,
\left(\frac{m_{3/2}}{2~\mathrm{TeV}}\right)\,
\left(\frac{m_\phi}{1500\,\mathrm{TeV}}\right)^{-1/2}
\,\left(\frac{M_\mathrm{P}}{M}\right)\;.
} & \:{\mbox{(B)}}
\end{array}
\right.
\nonumber
\end{eqnarray}
Cases A and B correspond to the model without and with the $\Phi^6$ term
in the superpotential, respectively.
By comparing the observed value $(n_b/s)^\mathrm{obs}=(8.7\pm0.3)\times10^{-11}$
\cite{Spergel:2006hy}, the lower bound on $m_\phi$ in Eq.~\eqref{eq:bound-mphi}
requires either enhanced SUSY breaking terms $|\kappa| \gg 1$ or the large
gravitino mass $m_{3/2} \gtrsim 10\,\mathrm{TeV}$ for case A whereas $m_{3/2}
\sim 1\,\mathrm{TeV}$ is possible in case B (assuming $M\simeq M_\mathrm{P}$).

As we shall see in the next section, the reference value of $m_\phi \sim
1500\,\mathrm{TeV}$ is motivated by considerations on dark matter. Notice that
the required size of the SUSY breaking scale $m_{3/2}$ from baryogenesis is
indeed at a favorable value for low energy phenomenology. In particular, in case
A, $m_{3/2} \sim 50\,\mathrm{TeV}$ indicates that loop corrections to soft
masses (anomaly mediation) are important. In many models with such situation,
the $W$ino~\cite{Giudice:1998xp,Randall:1998uk} or
Higgsino~\cite{Endo:2005uy,Choi:2005uz,Falkowski:2005ck} becomes the LSP and that is indeed
consistent with the discussion below. For case B, the $W$ino or Higgsino LSP is
realized by either assuming small value or phase of $\kappa^\prime$ such that
gravitino mass is enhanced or simply assuming the $W$ino or Higgsino LSP in the
scenarios of the gravity mediation type by relaxing the universality of the
gaugino and/or scalar masses.

\section{Dark matter from $\boldsymbol{\phi}$ decay}
\label{sec:DarkMatter}

Since every $\phi$ decay produces (at least) one superpartner,
Ref.~\cite{Thomas:1995ze} concludes that the number density of LSPs
exceeds the one of baryons, $n_\mathrm{LSP}\,\gtrsim\,n_b$. However,
$n_\mathrm{LSP}$ is modified by LSP pair annihilation processes in a
heavy $\phi$ scenario.  These processes are effective as long as the
corresponding rate exceeds the Hubble rate.
In the MSSM, the dark matter candidates which pair annihilate
strongly are the $W$ino and the Higgsino. Both particles have large
annihilation cross sections through weak interaction, and thus the
thermal abundance cannot explain the energy density of the dark
matter. On the other hand, non-thermal production from $\phi$ decays
renders the $W$ino/Higgsino a viable dark matter candidate.

In order to find out to what extent the LSPs annihilate, one describes the
evolution of number densities of $\phi$ quanta and LSPs, $n_\phi$ and $n_\chi$,
and energy density of the thermal bath, $\rho_\mathrm{rad}$, by Boltzmann
equations \cite{Moroi:1999zb},
\begin{subequations}\label{eq:Boltzmann}
\begin{eqnarray}
 \frac{\D n_\phi}{\D t} + 3H\,n_\phi
 & = & 
 -\Gamma_\phi\, n_\phi\;,\\
 \frac{\D n_\chi}{\D t} + 3H\,n_\chi
 & = & 
 \nu_\mathrm{LSP}\,\Gamma_\phi\, n_\phi-\langle\sigma\,v\rangle\,n_\chi^2\;,\\
 \frac{\D \rho_\mathrm{rad}}{\D t} + 4H\,\rho_\mathrm{rad}
 & = & 
 (m_\phi-\nu_\mathrm{LSP}\,m_\chi)\,\Gamma_\phi\, n_\phi
 +m_\chi \langle\sigma\,v\rangle\,n_\chi^2
 \;.
\end{eqnarray}
\end{subequations}
$\nu_\mathrm{LSP}$ denotes the number of LSPs produced by a $\phi$ decay. The
Boltzmann equations can be integrated, and the relic density of $\chi$ can be
approximated by \cite{Fujii:2001xp}
\begin{equation}\label{eq:nchiovers}
 \frac{n_\chi}{s}
 ~\sim~
 \left(4\,\langle\sigma v\rangle\, M_\mathrm{P}\,T_d\right)^{-1}\;,
\end{equation}
as long as the LSPs are not equilibrated, i.e.\ $T_d\lesssim m_\chi/30$. The
relic $\chi$ abundance is then
\begin{eqnarray}
 \Omega_\chi\,h^2
 & \simeq &
 0.1
\left(\frac{2.5\times10^{-3}}{m_\chi^2\langle\sigma v\rangle}\right)
\left(\frac{10^{-2}}{\xi}\right)^{1/2}
\left(\frac{m_\chi}{100\,\mathrm{GeV}}\right)^3\,
\left(\frac{m_\phi}{1500\,\mathrm{TeV}}\right)^{-3/2}\,
\left(\frac{M}{M_\mathrm{P}}\right)
 \;.\nonumber\\
 & &
 \label{eq:Relic$W$inoDensity}
\end{eqnarray}
The thermal average of the annihilation cross section is typically
$\langle \sigma v \rangle \sim 10^{-3} / m_\chi^2$ for the particles
which have $\mathrm{SU}(2)_\mathrm{L}$ quantum numbers such as $W$ino and
Higgsino. Therefore the non-thermal component can explain the dark
matter of the universe for $m_\phi \sim 10^{3-5}$~TeV depending on
$m_\chi$.

\begin{figure}[t]
\centerline{\CenterObject{\includegraphics{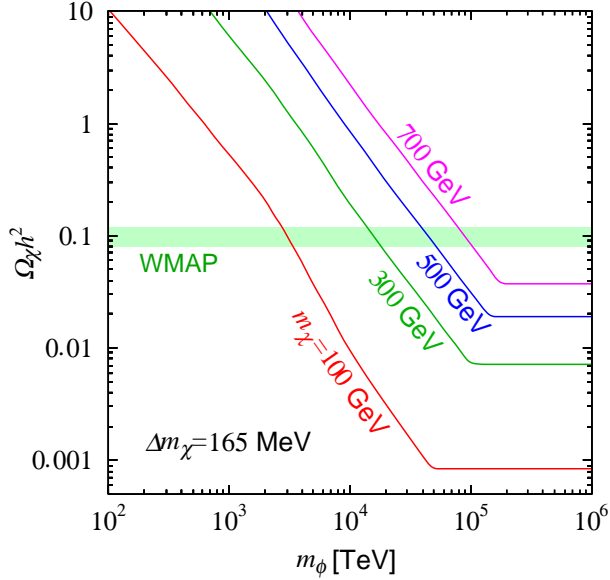}}}
\caption{$\Omega_\chi\,h^2$ as a function of the mass $m_\phi$ for
 various $W$ino masses $m_\chi$. We use for the mass difference between charged
 and neutral $W$inos $\Delta m_\chi=165\,\mathrm{MeV}$, and fix $M=M_\mathrm{P}$. 
 The shaded bar corresponds to the $2\sigma$ region of $\Omega_\mathrm{CDM}$ as
 reported by WMAP \cite{Spergel:2006hy}. 
As usual, we fix the present energy density of the universe so that
 $\Omega$ can formally become larger than one (`overclosure').}
 \label{fig:mphi_vs_Omega}
\end{figure}

For concreteness, let us focus on the case of the $W$ino LSP.
The annihilation cross section is \cite{Moroi:1999zb} (cf.\ the
extensive list \cite{Drees:1992am})
\begin{equation}\label{eq:SigmaMoroiRandall}
 \langle\sigma_{\widetilde{W}^0\widetilde{W}^0\to W^+ W^-} v\rangle
 \,=\,
 \frac{g_2^4}{2\pi}\frac{1}{m_{\chi}^2}
 \frac{\left[1-\frac{m_W^2}{m_\chi^2}\right]^{3/2}}{\left[2-\frac{m_W^2}{m_\chi^2}\right]^{2}}
 \;.
\end{equation}
In Fig.~\ref{fig:mphi_vs_Omega} we show the relic $W$ino density
$\Omega_\chi\,h^2$ (where $h\simeq0.7$ is the present normalized Hubble
expansion rate \cite{Spergel:2006hy}) as a function of $m_\phi$. To produce
Fig.~\ref{fig:mphi_vs_Omega}, we solve the set of Boltzmann equations
\eqref{eq:Boltzmann} (extended to include the charged $W$ino NLSP) and take into
account coannihilation. The coannihilation effect between LSP and NLSP becomes
important for large $m_\phi$, and explains the deviation of the contours from
straight lines (see in particular the $m_\chi = 100\,\mathrm{GeV}$ contour in
Fig.~\ref{fig:mphi_vs_Omega}). If $T_d\gtrsim m_\chi/30$, the $W$inos are in
thermal equilibrium, and the relic abundance does not depend on $m_\phi$ any
more. This explains why the contours become horizontal for large $m_\phi$ in
Fig.~\ref{fig:mphi_vs_Omega}.\footnote{For such large $m_\phi$, the standard
WIMP scenario works and thus the $B$ino is a good dark matter candidate.
However, this would require even larger $\phi$ masses, and according to
Eq.~(\ref{eq:baryon}), large $m_\phi$ would make baryogenesis difficult.
}
For $T_d\lesssim m_\chi/30$, the estimate \eqref{eq:Relic$W$inoDensity}
turns out to be a reasonable approximation. In particular, the
temperature $T_d$ can be as low as $100\,\mathrm{MeV}$ without
overclosing the universe. If we fix $M$ to be the Planck scale, we find
that $m_\phi$ should exceed $10^3\,\mathrm{TeV}$.

It is instructive to study the dependence of the baryon density in
Eq.~(\ref{eq:baryon})
and the relic dark matter density in Eq.~\eqref{eq:nchiovers} on the
physical parameters. For case A in Eq.~(\ref{eq:baryon}), amazingly,
the ratio is independent of the $\phi$ mass,
\begin{equation}
 \frac{n_\chi}{n_b}
 ~\sim~|\kappa|^{-1}\,
 \left( 4\,\langle\sigma v\rangle\,\xi\,m_{3/2}^2\right)^{-1}
 \,\left(\frac{M}{M_\mathrm{P}}\right)^2
 ~\sim~
 10^{4}\,|\kappa|^{-1}\,\left(\frac{m_\chi}{m_{3/2}}\right)^2
 \,\left(\frac{M}{M_\mathrm{P}}\right)^2
 \; .
\end{equation}
Our scenario relies on a large gravitino mass which is realized in anomaly
mediation \cite{Randall:1998uk,Giudice:1998xp} 
where the $W$ino mass is suppressed by a
loop factor, e.g.\ $m_\chi/m_{3/2}\sim g_2^2/(16\pi^2)$ for the $W$ino $\chi$.
This implies $n_\chi/n_b\sim (\text{few})\times 10^{-2}$ for $M\sim
M_\mathrm{P}$. Hence, our scenario predicts for the ratio of dark matter to
baryon densities
\begin{equation}
 \frac{\Omega_\chi}{\Omega_b}
 ~\sim~|\kappa|^{-1}\times
 \text{few}\times 10^{-2}\times\frac{m_\chi}{m_\mathrm{nucleon}}
 \times\left(\frac{M}{M_\mathrm{P}}\right)^2
 \;.
\end{equation} 
In particular, for $m_{\chi}$ of the order $100\,\mathrm{GeV}$ (and $M\simeq
M_\mathrm{P}$), the observed ratio $\Omega_\mathrm{CDM}/\Omega_b\simeq5$
\cite{Spergel:2006hy} finds a very natural explanation within the framework
described here.
The same is true for the Higgsino LSP case since it naturally has a mass
of the order of the $W$ino mass.

It is interesting to relax the assumption $M\simeq M_\mathrm{P}$ and
take, for instance, $M$ to be of order GUT or compactification scale,
$M\sim M_\mathrm{GUT}\simeq 3\times 10^{16}\,\mathrm{GeV}$, or the
string scale. If so, the $\phi$ mass can be substantially lower,
$m_\phi\sim m_{3/2}\sim100\,\mathrm{TeV}$.

Concerning cold dark matter, our analysis coincides with the one of
Ref.~\cite{Moroi:1999zb} if we identify $\phi$ as a modulus.
However, our assumption that $\phi$ is odd under $R$-parity completely
avoids the gravitino problem caused by the $\phi$ decay
\cite{Endo:2006zj,Nakamura:2006uc}.

\section{Conclusions}

We have discussed a scenario where both the observed baryon asymmetry
and the cold dark matter originate from decays of a heavy scalar field
$\phi$.
In an example we 
imposed $\mathbbm{Z}_{9}$ symmetry  such that $\phi$ number is effectively
conserved in the $\phi$ oscillation era. This allows to have a initial asymmetry
$q_\phi$ to be conserved until the $\phi$ decay, and $q_\phi$ is converted into
baryon asymmetry. 
The baryon asymmetry is automatically in the right ballpark for $\phi$ masses
which are high enough to evade the BBN constraints.
The $\phi$ decays also produce LSPs.  For sufficiently high decay
temperature $T_d$, pair annihilation is still partially effective if the
LSP is the $W$ino or the Higgsino, and hence the number density of LSPs
gets reduced into a correct size for the dark matter, $n_\chi\ll n_b$.
In one of our scenarios (case A) we require a heavy gravitino,
$m_{3/2}\sim100\,\mathrm{TeV}$, loop contributions to gaugino masses become
important, so that one naturally obtains a $W$ino or Higgsino LSP.

Amazingly, in that context, our mechanism predicts $\Omega_\chi/\Omega_b
\,\sim\,\text{few}\times 10^{-2}\times(m_\chi/m_\mathrm{nucleon})$
independently of the $\phi$ mass $m_\phi$.
It can hence very naturally account for $\Omega_\mathrm{\chi}\simeq
5\,\Omega_b$ although each $\phi$ decay produces (at least) one
superpartner and $m_\mathrm{LSP}\gtrsim100\times m_\mathrm{nucleon}$. In
our scenario, the role of the $\phi$ field is thus twofold: it serves
both as source of the observed baryon asymmetry and explains why $W$inos
or Higgsinos can be cold dark matter despite of the large annihilation
cross section.

Our mechanism also opens new possibilities in inflation model building. One
could, for instance, envisage a sequestered scenario where the inflaton is
(geometrically) separated from the MSSM fields. This is usually a problem
because then the inflaton reheats the hidden sector. In our scenario, however,
such a setup may be viable because (dark) matter and radiation are generated by
the $\phi$ field rather than inflaton decays.

\subsubsection*{Acknowledgements}

We would like to thank K.~Choi, K.~Hamaguchi, O.~Lebedev and H.P.~Nilles
for useful discussions.  
We acknowledge support from the Aspen Center
for Physics where this work has been started.
The research of MR is supported by the DFG cluster of excellence Origin
and Structure of the Universe, the Graduiertenkolleg ``Particle Physics at the
Energy Frontier of New Phenomena'' and the SFB-Transregios 27 ``Neutrinos and
Beyond'' by the Deutsche Forschungsgemeinschaft (DFG).  The research of
HM is supported in part by World Premier International Research Center
Initiative (WPI Initiative), MEXT, Japan, in part by the
U.S. Department of Energy under Contract DE-AC03-76SF00098, and in
part by the National Science Foundation under grant PHY-04-57315. 

\bibliography{BAU}
\addcontentsline{toc}{section}{Bibliography}
\bibliographystyle{NewArXiv}

\end{document}